\documentclass[aps,prl,showpacs,10pt, superscriptaddress,twocolumn,byrevtex,floatfix]{revtex4-1}
\usepackage{graphicx}
\usepackage{amsmath}
\usepackage{braket}
\usepackage{color}
\usepackage{hyperref}
\usepackage{tikz}
\usepackage{array}
\usepackage[section]{placeins}

\hypersetup{bookmarksnumbered,%
	colorlinks,%
	linkcolor=blue,%
	citecolor=blue,%
	plainpages=false,%
	pdfstartview=FitH}

\newcommand{\nc}{\newcommand}           % new command
\nc{\vc}[1]     {\mbox{\boldmath $#1$}} % boldmath(bmtor)
\nc{\mapleft}[1]{                       % something under arrow
 \smash{\mathop{                        %
  \hbox to 0.90cm{\rightarrowfill} }\limits_{#1}}}
\nc{\figwidth}{0.55}                    % figure width, in unit of textwidth, preprint
\nc{\mydraft}	{\setlength{\topmargin}{-1.5cm}}
\mydraft

\begin{document}
%\preprint{}

\title{Hypernuclear cluster states of $_\Lambda^{12}\rm{B}$ Unveiled through Neural Network-Driven Microscopic Calculation}

\author{Jiaqi Tian}
\affiliation{College of Physics, Nanjing University of Aeronautics and Astronautics, Nanjing 210016, China}
\affiliation{Key Laboratory of Aerospace Information Materials and Physics (NUAA), MIIT, Nanjing 211106, China}
\author{Mengjiao Lyu}
\email{mengjiao.lyu@nuaa.edu.cn}
\affiliation{College of Physics, Nanjing University of Aeronautics and Astronautics, Nanjing 210016, China}
\affiliation{Key Laboratory of Aerospace Information Materials and Physics (NUAA), MIIT, Nanjing 211106, China}

\author{Zheng Cheng}
\affiliation{College of Physics, Nanjing University of Aeronautics and Astronautics, Nanjing 210016, China}
\affiliation{Key Laboratory of Aerospace Information Materials and Physics (NUAA), MIIT, Nanjing 211106, China}

\author{Masahiro Isaka}
\affiliation{Hosei University, 2-17-1 Fujimi, Chiyoda-ku, Tokyo 102-8160, Japan}

\author{Akinobu Doté}
\affiliation{KEK Theory Center, Institute of Particle and Nuclear Studies (IPNS), High Energy Accelerator Research Organization (KEK), 1-1 Oho, Tsukuba, Ibaraki, 305-0801, Japan}
\affiliation{J-PARC Branch, KEK Theory Center, IPNS, KEK, 203-1, Shirakata, Tokai, Ibaraki, 319-1106, Japan}

\author{Takayuki Myo}
\affiliation{Research Center for Nuclear Physics (RCNP), Osaka University, Osaka 567-0047, Japan}
\affiliation{General Education, Faculty of Engineering, Osaka Institute of Technology, Osaka, Osaka 535-8585, Japan}

\author{Hisashi Horiuchi}
\affiliation{Research Center for Nuclear Physics (RCNP), Osaka University, Osaka 567-0047, Japan}

\author{Hiroki Takemoto}
\affiliation{Faculty of Pharmacy, Osaka Medical and Pharmaceutical University, Takatsuki, Osaka 569-1094, Japan}

\author{Niu Wan}
\affiliation{School of Physics and Optoelectronics, South China University of Technology, Guangzhou 510641, China}

\author{Qing Zhao}
\affiliation{School of Science, Huzhou University, Huzhou 313000, Zhejiang, China}

\date{\today}


\date{\today}

\begin{abstract}
  We investigate the hypernuclear cluster states of $_\Lambda^{12}\mathrm{B}$ using a neural-network-driven microscopic model. We extend the Control Neural Networks (Ctrl.NN) method and systematically calculate the positive-parity spectrum of $_\Lambda^{12}\mathrm{B}$. By incorporating $sd$-shell excitations and parity-coupling effects into the $_\Lambda^{12}\mathrm{B}$ hypernuclear system, we reveal structural changes, including clustering effects and new configurations such as isosceles-triangle and $\alpha$-$t$-$\alpha$ linear-chain structures. Furthermore, by comparing with experimental data, we identify that many peaks ($\#$6 and $\#$8) can be interpreted as $p_\Lambda$ dominant states, which is consistent with shell-model predictions. Notably, based on our analysis of the excited states of $_\Lambda^{12}\mathrm{B}$, we propose possible candidates for previously unexplained or controversial experimental peaks.
\end{abstract}
\maketitle

%\keywords{}

\section{Introduction}
In $\Lambda$ hypernuclei, the ``strangeness" degree of freedom gives rise to various intriguing phenomena through the introduction of a $\Lambda$ hyperon, which contains one strange quark. 
Thus, one of the primary goals of hypernuclear physics is to understand the baryonic many-body system, beginning with the hyperon-nucleon ($YN$) and nucleon-nucleon ($NN$) interactions \cite{hiyama_structure_2009}. 
Among these phenomena, an intriguing finding is the ``glue-like role" of the $\Lambda$ hyperon, which induces dynamic changes in the core nucleus, as evidenced by the electric transition $B(E2)$ \cite{motoba_progr_1983, tanida_measurement_2001}. 
For instance, in the $^{7}_\Lambda\mathrm{Li}$ hypernucleus, the experimental measurement of $B(E2)$ reveals significant shrinkage of the system due to the addition of the $\Lambda$ hyperon, compared with the conventional core nucleus \cite{tanida_measurement_2001}. 
Moreover, when an additional $\Lambda$ hyperon is introduced, the well-developed cluster states undergo even more pronounced shrinkage than compact, shell-like states, thereby explaining the notable differences in their $\Lambda$ binding energies. In other words, the $\Lambda$ binding energy is highly sensitive to the structure of the core nucleus. 
For example, the $\Lambda$ binding energies of $^{12}\mathrm{C}(0_2^+) \otimes s_\Lambda$ \cite{hiyama_three-body_1996,mei_microscopic_2015} and $^{10}\mathrm{Be}(0_2^+) \otimes s_\Lambda$ \cite{isaka_impurity_2015} are predicted to be about 3 MeV lower than those of the corresponding ground states. 
On the other hand, the $^{12}\mathrm{C}(0_2^+)$ state is well known for its gas-like cluster structure \cite{Tohsaki2001}, while in $^{10}\mathrm{Be}(0_2^+)$, two clusters and valence nucleons form a linear-like $\sigma$-molecular orbital \cite{Itagaki2000, Lyu2016}, suggesting a considerable correlation between clustering and the $\Lambda$ binding energy. 
In the spectra of the corresponding hypernuclei, differences in $B_\Lambda$ can be seen through level positions and gaps, relative to the conventional core nucleus.

Experimentally, owing to a modified setup, the large electromagnetic background was significantly suppressed, enabling hypernuclear spectroscopy with higher energy resolution using the $(e, e'K)$ reaction in $^{12}_\Lambda \mathrm{B}$ \cite{iodice_high_2007, tang_experiments_2014}, $^{10}_\Lambda \mathrm{Be}$ \cite{gogami_high_2016}, and $^{9}_\Lambda \mathrm{Li}$ \cite{gogami_spectroscopy_2021} at JLab.
Notably, the spectrum of $^{12}_\Lambda \mathrm{B}$ was measured in $(e, e'K)$ experiments at JLab Hall A \cite{iodice_high_2007} and Hall C \cite{tang_experiments_2014}, yielding several experimental peaks and corresponding cross sections in the excitation-energy range of $E_x=0-12$ MeV.
Some peaks near the $^{11}\mathrm{B}+\Lambda$ threshold are interpreted as $p$-wave excitations of the $\Lambda$ hyperon based on a 0$\hbar\omega$ shell-model calculation \cite{iodice_high_2007}, although peaks $\#$4 and $\#$7 remain unexplained.
However, in the $sd$-shell excitation of the $^{11}\mathrm{B}$ core, many positive-parity states appear in the same energy region \cite{yamada__2010, suhara_cluster_2012, zhou_2_2018}.
Many of these states are considered to exhibit varying degrees of clustering \cite{zhou_2_2018}.
Hence, there are two types of excitation modes: (1) core-excited states $\otimes s_\Lambda$ and (2) low-lying core states $\otimes p_\Lambda$.
These two excitation modes can couple to form states with unnatural parity of the core nucleus ($\Lambda$ particle), referred to as the parity coupling (or intershell coupling) effect \cite{Yamada1984NeHypernucleus, motoba_parity-mixing_1998}.
Therefore, the unexplained peaks and the significant structural changes of cluster states arouse our
interest in reconsidering this system by taking account of the $sd$-shell excitation and parity coupling effect.

To fully describe the nature of clustering in core-excited states as well as the parity coupling effect, we adopt the microscopic cluster model to investigate the $^{12}_\Lambda \rm{B}$ hypernucleus.
We expect to explain the experimental results and provide new insight into the $^{12}_\Lambda \rm{B}$ system through the microscopic cluster model calculations.
The paper is organized as follows. In Section.~II, the theoretical framework of the Hyper-Brink wave function for the $^{12}_\Lambda \rm{B}$ hypernucleusare is explained. In Sec.~III, 
we present the numerical results and discuss the structural properties of $^{12}_\Lambda \rm{B}$ hypernucleus. 
Finally, in Sec.~IV, we provide our conclusion.

\section{THEORETICAL FRAMEWORK}
\label{sec:frame}
We utilize the multi-cooling method \cite{myo_variation_2023, Myo2025}, guided by the Control Neural Network (Ctrl.NN), to optimize the Hyper-Brink wave function, following the approach in Ref.~\cite{tian_lambda_2024}.
The Ctrl.NN is a powerful method for optimizing the total Hyper-Brink wave function and investigating hypernuclear structures.
In this Letter, we further update this framework by imposing the $K$-projection \cite{shikata_variation_2020} on the variational multi-basis wave function, where the Hyper-Brink bases are projected onto the definite $K^\pi$.
Detailed descriptions of the Ctrl.NN can be found in Ref.~\cite{tian_lambda_2024,cheng_evidence_2024}.
\subsection{Hyper-Brink wave function}
In this work, we employ a microscopic 2$\alpha+t+\Lambda$ four-body cluster model to elucidate the cluster properties of hypernuclear structures.
The single Hyper-Brink wave function is fully antisymmetrized, with the single-particle wave functions represented by Gaussian forms \cite{tian_lambda_2024}.
\begin{equation}\label{eq:amd-k}
  \Phi_B = 
    \phi^{\Lambda}(\boldsymbol{R}_\Lambda, s_\Lambda)\mathcal{A}\left\{
      \Phi_\alpha(\boldsymbol{R}_{1})
      \Phi_\alpha(\boldsymbol{R}_{2})
      \Phi_t(\boldsymbol{R}_{t}, s_t)
    \right\},
\end{equation}
where
\begin{equation}
  \Phi_\alpha(\boldsymbol{R}) = 
    \mathcal{A}\left\{
      \phi(\boldsymbol{R})\chi_{p\uparrow}
      \phi(\boldsymbol{R})\chi_{p\downarrow}
      \phi(\boldsymbol{R})\chi_{n\uparrow}
      \phi(\boldsymbol{R})\chi_{n\downarrow}
    \right\},
\end{equation}
\begin{equation}
  \Phi_t(\boldsymbol{R}_t, s) = 
    \mathcal{A}\left\{
      \phi(\boldsymbol{R}_t)\chi_{ps}
      \phi(\boldsymbol{R}_t)\chi_{n\uparrow}
      \phi(\boldsymbol{R}_t)\chi_{n\downarrow}
    \right\}.
\end{equation}
The $\Phi_\alpha$ and $\Phi_t$ represent the wave functions of $\alpha$ and triton clusters. 
The $\phi^{\Lambda}$ and $\phi^{N}$ are single-particle wave functions of $\Lambda$ particle and nucleons, respectively, with width parameters $\nu_{\Lambda}$ and $\nu_{N}$, as
\begin{equation}
  \begin{aligned}
  \phi^{\Lambda}(\boldsymbol{R}_\Lambda, s_\Lambda)&=
      \left(\frac{2\nu_{\Lambda}}{\pi}\right)^{3/4}
      \exp\left\{
          -\nu_{\Lambda}(\boldsymbol{r}_\Lambda-\boldsymbol{R}_\Lambda)^{2}
      \right\}\chi_{s_\Lambda},
  \end{aligned}
\end{equation}
\begin{equation}
  \begin{aligned}
  \phi^{N}(\boldsymbol{R})&=
      \left(\frac{2\nu_{N}}{\pi}\right)^{3/4}
      \exp\left\{
          -\nu_{N}(\boldsymbol{r}-\boldsymbol{R})^{2}
      \right\}.
  \end{aligned}
\end{equation}
We adopt $\nu_{N} = 0.222 , \rm{fm}^{-2}$, consistent with Ref.~\cite{zhou_2_2018}, and set $\nu_{\Lambda} = \nu_{N}(M_{\Lambda}/M_{N})^2$, where $M_{\Lambda}$ and $M_{N}$ are the masses of the $\Lambda$ particle and the nucleon.
The $\boldsymbol{R}_\Lambda$ and $\boldsymbol{R}$ are the generator coordinates for the $\Lambda$ particle and nucleons, determining the spatial configuration of the single Hyper-Brink wave function.
$\chi_{p\uparrow}\cdots\chi_{n\downarrow}$ denote the spin-isospin components of the nucleon wave functions, while $\chi_{s_\Lambda}$ represents the spin wave function of the $\Lambda$ particle.
In the triton cluster wave function, the spin of the proton is represented by $s$, which can be either up or down.

\subsection{Variational wave function}
During the variation (multi-cool) process \cite{myo_variation_2023} guided by the Control Neural Network \cite{tian_lambda_2024,cheng_evidence_2024}, we perform the $K$- and parity-projection (K-VAP) for single Hyper-Brink basis \cite{shikata_variation_2020}.
The projected Hyper-Brink bases are constructed through the antisymmetrization of single-particle wave functions, as follows:
\begin{equation}\label{eq:amd-k}
  \Phi_B^{K;\pm}=\frac{1}{\sqrt{n!}}\hat{P}^K\hat{P}^\pm
    \Phi_B,
\end{equation}
where $\mathcal{A}$ and $\hat{P}^\pm$ are the antisymmetrization operator and parity projection operator respectively. $n$ is the number of nucleons. $\hat{P}^K$ is the $K$-projection operator, which can be given as
\begin{equation}
    \hat{P}^K=\frac{1}{2 \pi} \int_0^{2 \pi} d \theta e^{-i K \theta} \hat{R}(\theta),
\end{equation}
where the $\hat{R}(\theta)$ is the rotation operator around the principal axis in a body-fixed frame.
After this projection, the Hyper-Brink basis can be optimized for a definite $K^\pi$. 
By recovering rotational symmetry around the principal axis, the Ctrl.NN method explores the effective model space rather than restoring rotational symmetry during the multi-cooling.
The latter approach attempts to reconstruct the total wave function with rotational symmetry, resulting in minor discrepancies among the bases after the total angular momentum projection ($J$-projection).
Although both approaches reduce the energy (the loss function of Ctrl.NN), the latter is less effective because we diagonalize the updated basis states after the total angular momentum projection in the final calculation. 
Hence, the K-VAP method significantly reduces the model space and enhances the efficiency of the Hyper-Brink basis.
As shown in Ref.~\cite{shikata_variation_2020}, the AMD wave function is further optimized after the $K$-projection.

Another critical aspect is the spin constraint applied during the variation process.
Due to the strong correlation between spin configurations and hypernuclear states, all possible spin configurations should be considered in the present calculation.
In this study, we constrain the spins of the $\Lambda$ particle and the proton in the triton cluster to four combinations: $\Lambda^\uparrow p^\uparrow$, $\Lambda^\uparrow p^\downarrow$, $\Lambda^\downarrow p^\uparrow$, and $\Lambda^\downarrow p^\downarrow$.
Under the spin constraint, we obtain basis states with diverse spatial configurations during the multi-cooling process, thereby covering a broader model space.

Next, we define the total wave function during the multi-cooling, which is the linear superposition of the Hyper-Brink basis.
\begin{equation}\label{total_wave_function}
  \begin{aligned}
    |\Psi^{K;\pm}\rangle=\sum_{i=1}^m C_i\left|\Phi^{K;\pm(i)}_B\left(\boldsymbol{R}_{\Lambda}^{(i)}, \boldsymbol{R}_{ 1}^{(i)}, \boldsymbol{R}_{ 2}^{(i)},  \boldsymbol{R}_{ t}^{(i)}\right)\right\rangle \text {, }  
  \end{aligned}
\end{equation}
where $C_i$ are the superposition coefficients of each Hyper-Brink basis. The variational parameters, serving as both input and output in Ctrl.NN, are all the generator coordinates, as
\begin{equation}
  \begin{aligned}
  \left\{\boldsymbol{R}\right\}^{(i)}  =\boldsymbol{R}_{\Lambda}^{(i)}, \boldsymbol{R}_{1}^{(i)},  \cdots, \boldsymbol{R}_{n}^{(i)}.
  \end{aligned}
\end{equation}
The coefficients $C_i$ and the energy eigenvalue $E$ are determined by solving the eigenvalue problem.
The output parameters are then used to optimize the neural network, as described in our previous work \cite{tian_lambda_2024}.
The Ctrl.NN optimize multiple sets of basis states and learn the properties of the
quantum states with different 
$K$ quantum numbers and spin couples. 
Finally, we combine all these basis states and solve the eigenvalue problem after applying the total angular momentum projection ($J$-projection). 
\subsection{Hamiltonian}
The Hamiltonian used in present calculation is composed of the kinetic energy of $\Lambda$ hyperon ($T_\Lambda$) and nucleon ($T_N$), 
nucleon-nucleon central interaction $V_{i j}^{(N N)}$, $\Lambda N$ central interaction $V_i^{(\Lambda N)}$, 
nucleon-nucleon spin-orbit coupling interaction $V_{i j}^{(N N l s)}$ and Coulomb interaction $V_{i j}^{(C)}$,
\begin{equation}
    \begin{aligned}
    H = &\sum_{i=1}^{11} T_i^N+T^{\Lambda}-T_G+\sum_{i<j}^{11} V_{i j}^{(C)}\\
    & +\sum_{i<j}^{11} V_{i j}^{(N N)}+\sum_{i<j}^{11} V_{i j}^{(N N l s)}+\sum_{i=1}^{11} V_i^{(\Lambda N)}.
    \end{aligned}
\end{equation}
We eliminate the effects of spurious center-of-mass motion, denoted as $T_G$, in the Hamiltonian. 
In this Letter, we focus on states that couple the $sd$-shell excitation of core nucleus and $\Lambda$ hyperon which occupies $s$-wave. 
In this Letter, we focus on states that couple the $sd$-shell excitations of the core nucleus with a $\Lambda$ hyperon occupying the $s$-wave.
Hence, the $\Lambda N$ spin-orbit coupling interaction is sufficiently weak that we neglect it. 
We adopt Volkov No.2 nucleon-nucleon central force $V_{i j}^{(N N)}$ and set the Majorana parameter $M = 0.598$. 
The G3RS interaction is used for the nucleon-nucleon spin-orbit coupling interaction $V_{i j}^{(N N l s)}$. The strength of G3RS interaction is adopted as 1600 MeV.
The nucleon-nucleon interaction parameters are same as adopted Set 2 of parameters in Ref.~\cite{zhou_2_2018}, where the excited energy $E_x$ of $sd$-shell excited states can be reproduced.

For $\Lambda N$ force $V_i^{(\Lambda N)}$, we adopt ESC14 interaction with many body effect (MBE), which is an effective local interaction with the Gaussian form by simulating the G-matrices calculation from the Nijmegen ESC potential, where the $\Lambda N-\Sigma N$ coupling is renormalized by the G-matrix calculation. In this paper, we adopted the tuning version proposed in Ref.~\cite{isaka_low-lying_2020}, where the spin-dependent part is tuned to reproduce the splitting between $1/2^+$ and $3/2^+$ states of $^{7}_\Lambda\rm{Li}$.

The $\Lambda$N$G$-matrix interaction $V_{\Lambda N}^{\text {cent }}$ is written as
\begin{equation}
    \begin{aligned}
        V_{\Lambda N}^{\text {cent }}= & v^{\left({ }^{1} E\right)} \hat{P}\left({ }^{1} E\right)+v^{\left({ }^{3} E\right)} \hat{P}\left({ }^{3} E\right) \\
        &+v^{\left({ }^{1} O\right)} \hat{P}\left({ }^{1} O\right)+v^{\left({ }^{3} O\right)} \hat{P}\left({ }^{3} O\right)
    \end{aligned}
\end{equation}
where 
\begin{equation}
\begin{aligned}
    v^{(c)}\left(k_{F}, r\right) &=\sum_{i=1}^{3}\left(a_{i}^{(c)}+b_{i}^{(c)} k_{F}+c_{i}^{(c)} k_{F}^{2}\right) e^{-r^{2} / \beta_{i}^{2}},
\end{aligned}
\end{equation}
and
\begin{equation}
\begin{aligned}
    c &={ }^{1} E,{ }^{3} E,{ }^{1} O \text { or }{ }^{3} O.
\end{aligned}
\end{equation}
The parameter $a^{(c)}_i$, $b^{(c)}_i$, $c^{(c)}_i$, and $\beta_i$ in four channels of ESC14 are listed in Ref.~\cite{isaka_low-lying_2020}.
Using Pauli's spin matrix $\bm{\sigma}$, $\Lambda$-N force $V_i^{(\Lambda N)}$ can be rewritten as
\begin{equation}
\begin{aligned}
    V_{\Lambda N}^{\text {cent }}=& \sum_{i=1}^{3}\left\{\left(v_{0}^{E(i)}+v_{\sigma}^{E(i)} \bm{\sigma} \cdot \bm{\sigma}\right) \hat{P}(E)\right.\\
    &\left.+\left(v_{0}^{O(i)}+v_{\sigma}^{O(i)} \bm{\sigma} \cdot \bm{\sigma}\right) \hat{P}(O)\right\} e^{-r^{2} / \beta_{i}^{2}}.
\end{aligned}
\end{equation}
As shown in Ref.~\cite{isaka_low-lying_2020}, the $\Lambda$ binding energies of various 
$\Lambda$ hypernuclei can be systematically reproduced by employing the 
HyperAMD calculation with ESC14+MBE interaction.
In the present calculation of $_\Lambda^{12}\rm{B}$,  we adopt the Fermi momentum parameter $k_F$ of 1.06 fm$^{-1}$ to clarify the excitation of the core nucleus and $\Lambda$ hyperon, which well reproduces the $\Lambda$ binding energy $B_\Lambda$ of the ground state as 11.53 MeV.

\section{results and discussions}
\label{sec:results1}
\begin{figure}[htbp] %H为当前位置，!htb为忽略美学标准，htbp为浮动图形
  \centering %图片居中
  \includegraphics[width=0.48\textwidth]{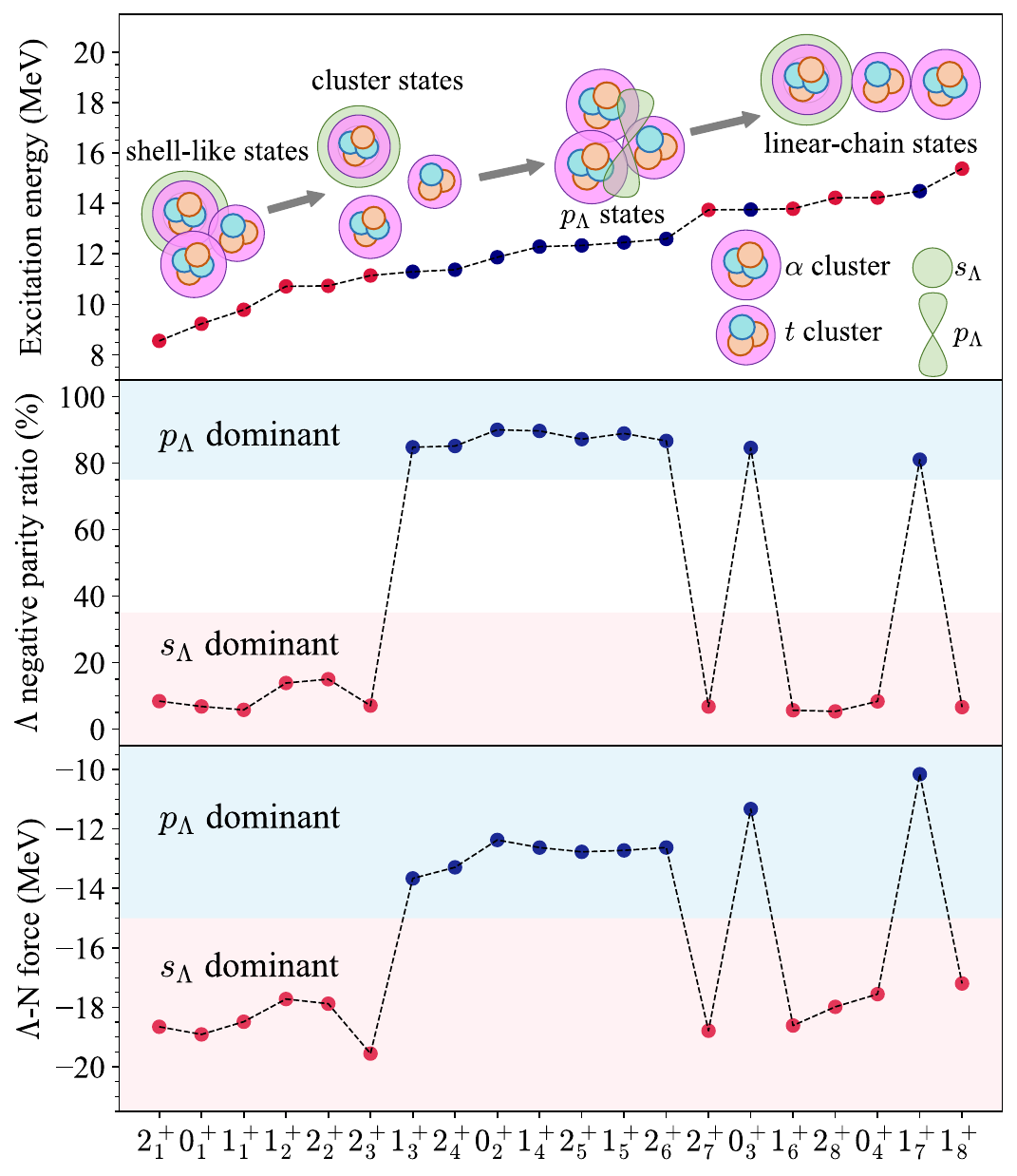} %插入图片，[]中设置图片大小，{}中是图片文件名
  \caption{The excitation energies, $\Lambda$ negative-parity components, and $\Lambda$-N force of all states are shown. The $\Lambda$ negative-parity component is explained in the Appendix. The structural evolution of $_\Lambda^{12}\rm{B}$ positive-parity states is described with diagrammatic representation. The blue and pink regions in the middle and bottom figures separate the different single-particle orbit of $\Lambda$ hyperon.} %最终文档中希望显示的图片标题
  \label{LNforce} %用于文内引用的标签
\end{figure}

Before investigating the $_\Lambda^{12}\rm{B}$, we first calculate the positive-parity excited states of core nucleus $^{11}\rm{B}$ in Table~\ref{tab} and briefly summarize their structures. 
Overall, the excitation energies of these states are reproduced although the $1/2^+_2$ and $3/2^+_2$ states are slightly overestimated. 
Most of these states exhibit developed clustering, with a large inter-cluster distance between two $\alpha$ clusters and a broad distribution of the $t$ cluster \cite{zhou_2_2018}. 
However, among these states, the $5/2_1^+$ state exhibits a relatively compact shell-like structure. On the other hand, some states, such as $1/2_2^+$ and $5/2_3^+$, exhibit the linear-chain configurations, which can also be explained as the $(\sigma)^3$ orbital in the molecular orbit picture within the AMD framework \cite{suhara_cluster_2012}.
The pronounced clustering nature of these states encourages us to study their properties by employing the cluster model, ensuring reasonable analysis of the results.
\begin{table}[htbp]
  \centering
  \renewcommand{\arraystretch}{1.5} 
  \caption{The excitation energies for each state in the core nuclei $^{11}\rm{B}$ in a unit of MeV. The experimental values reported in Ref~\cite{kelley2012energy} are also presented. The experimental value of $1/2^+_2$ with the uncertain identification is enclosed in the parenthesis.}
  \label{tab}
  \setlength{\tabcolsep}{0.8mm}
  {\begin{tabular}{ccccccccccc}
    \hline\hline 
    $J^\pi$ & $E_x$ & Expt & &$J^\pi$ & $E_x$ & Expt & &$J^\pi$ & $E_x$ & Expt \\
    \cline{1-3} \cline{5-7} \cline{9-11} 
    $1/2^-_1$ & 0.74  & 2.12 & &$3/2^-_1$ & 0.0 & 0.0 & &$5/2^+_1$ & 6.44 & 7.28 \\
    $1/2^+_1$ & 6.24  & 6.79 & &$3/2^+_1$ & 7.61 & 7.98 & &$5/2^+_2$ & 9.19 & 9.27 \\
    $1/2^+_2$ & 11.44  & (9.82) & &$3/2^+_2$ & 11.16 & 9.87 & &$5/2^+_3$ & 11.09 & 11.60 \\
    % ($1/2^+$)& -----  & 12.55 & & & &  & &$5/2^+_3$ & 11.09 & 11.60 \\
    \hline\hline
  \end{tabular}}
\end{table}
% % During the multi-cooling process, we fix the spins of the triton and $\Lambda$ hyperon in four combinations: $\Lambda^\uparrow$ $t^\uparrow$, $\Lambda^\uparrow$ $t^\downarrow$, $\Lambda^\downarrow$ $t^\uparrow$, $\Lambda^\downarrow$ $t^\downarrow$.
% Due to the existence of spin-dependent terms in the NN and YN interactions, the multi-basis system with different spin combinations evolves into different model spaces as time develops.
\begin{figure*}[htbp] %H为当前位置，!htb为忽略美学标准，htbp为浮动图形
  \centering %图片居中
  \includegraphics[width=0.78\textwidth]{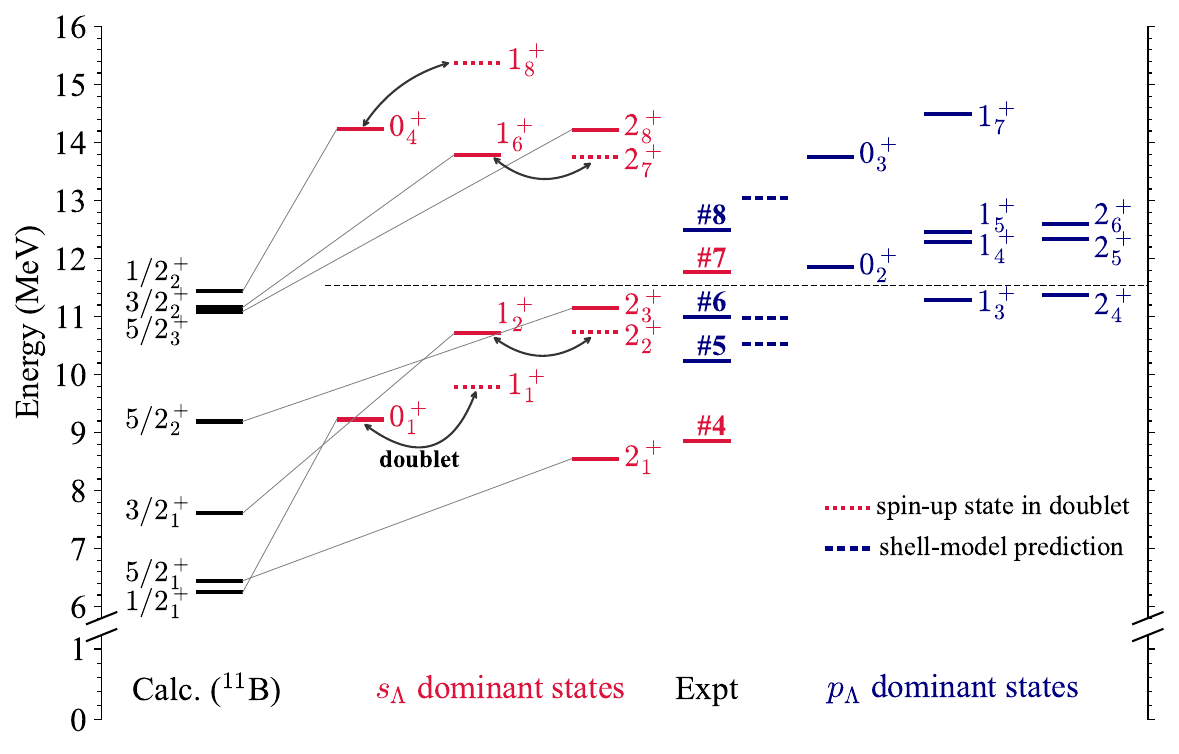} %插入图片，[]中设置图片大小，{}中是图片文件名
  \caption{Comparison of the excitation energies of $_\Lambda^{12}\rm{B}$ together with those of the core nucleus $^{11}$B from present calculations with experiments \cite{tang_experiments_2014}. The shell-model prediction \cite{iodice_high_2007} is also shown beside the experimental data with blue dashed lines. The black dashed line represents the $^{11}$B+$\Lambda$ threshold} %最终文档中希望显示的图片标题
  \label{spectrum} %用于文内引用的标签
\end{figure*}
In Fig.~\ref{spectrum}, many positive-parity states of $_\Lambda^{12}\rm{B}$ are obtained by adding a $\Lambda$ hyperon.
The shell-model prediction \cite{iodice_high_2007} is also shown for comparison with present calculations. 
Due to the introduction of the $\Lambda$ hyperon and account of parity coupling, numerous states appear around the $\Lambda$ breakup threshold.
Hence, the first challenge is to clearly identify the excitation modes of each state.
We define $s_\Lambda$ ($p_\Lambda$) dominant states, where the $\Lambda$ hyperon occupies the $s$ orbit ($p$ orbit). 
Considering the possibility of parity coupling, we perform two calculations to distinguish the single-particle orbit of the $\Lambda$ hyperon and evaluate the coupling levels for each state of $_\Lambda^{12}\rm{B}$.
First, we calculate the $\Lambda$-N force for all states because the interaction between the hyperon and nucleons is weaker when the hyperon occupies a higher orbit.
Fig.~\ref{LNforce} shows the variation of the $\Lambda$-N force with the increase of excitation energy.
It shows a distinct excitation energy region around 12 MeV, indicating that the $\Lambda$ hyperon is excited to the $p$ orbit.

Second, we calculate the negative-parity component of $\Lambda$ hyperon, which suggests the $\Lambda$ hyperon has been excited from $s$ orbit and the different degrees of parity coupling emerge in all of the states (see the Appendix).
As a benchmark for the excited states, the $\Lambda$ negative-parity component of the ground state is just 3$\%$.
In positive-parity excited states, the $s_\Lambda$ dominant states and $p_\Lambda$ dominant states perform around 10$\%$ and 85$\%$ negative-parity components of $\Lambda$, respectively.
The results suggest that $s_\Lambda$ and $p_\Lambda$ couple in certain excited states to form unnatural parities and different coupling levels are clearly shown.
Additionally, by comparing the Hamiltonian components and structures of $s_\Lambda$ dominant states, we identified four pairs of doublet states: ($0^+_1\leftrightarrow1^+_1$), ($0^+_4\leftrightarrow1^+_8$), ($1^+_2\leftrightarrow2^+_2$), and ($1^+_6\leftrightarrow2^+_7$).
In Fig. \ref{spectrum}, these doublet states are shown, where the states with $J+1/2$ are exhibited by dotted lines to clarify the full spectrum.
On account of the identification mentioned above, the states of  $_\Lambda^{12}\rm{B}$ conveniently correspond to those of core nucleus states.

In the calculated spectrum of $_\Lambda^{12}\rm{B}$, the experimental peaks $\#$6 and $\#$8 are well reproduced as the $p_\Lambda$ dominant states by comparing the present calculation, which is consistent with the shell-model prediction.
On the other hand, we also predict many of $s_\Lambda$ dominant states. The introduction of $s_\Lambda$ significantly increases the excitation energies because of the weaker attraction between $\Lambda$ hyperon and nucleons in dilute cluster states. 
Furthermore, due to the varying degrees of clustering in $3/2_1^+$ and $5/2_2^+$, the level spacing is also changed. 
Notably, the inversion of level ordering occurs between $5/2_1^+$ and $1/2_1^+$ as expected, attributed to the dynamic difference between the compact shell-like structure of $5/2_1^+$ and pronounced clustering of $1/2_1^+$.

\begin{figure}[htbp] %H为当前位置，!htb为忽略美学标准，htbp为浮动图形
  \centering %图片居中
  \includegraphics[width=0.45\textwidth]{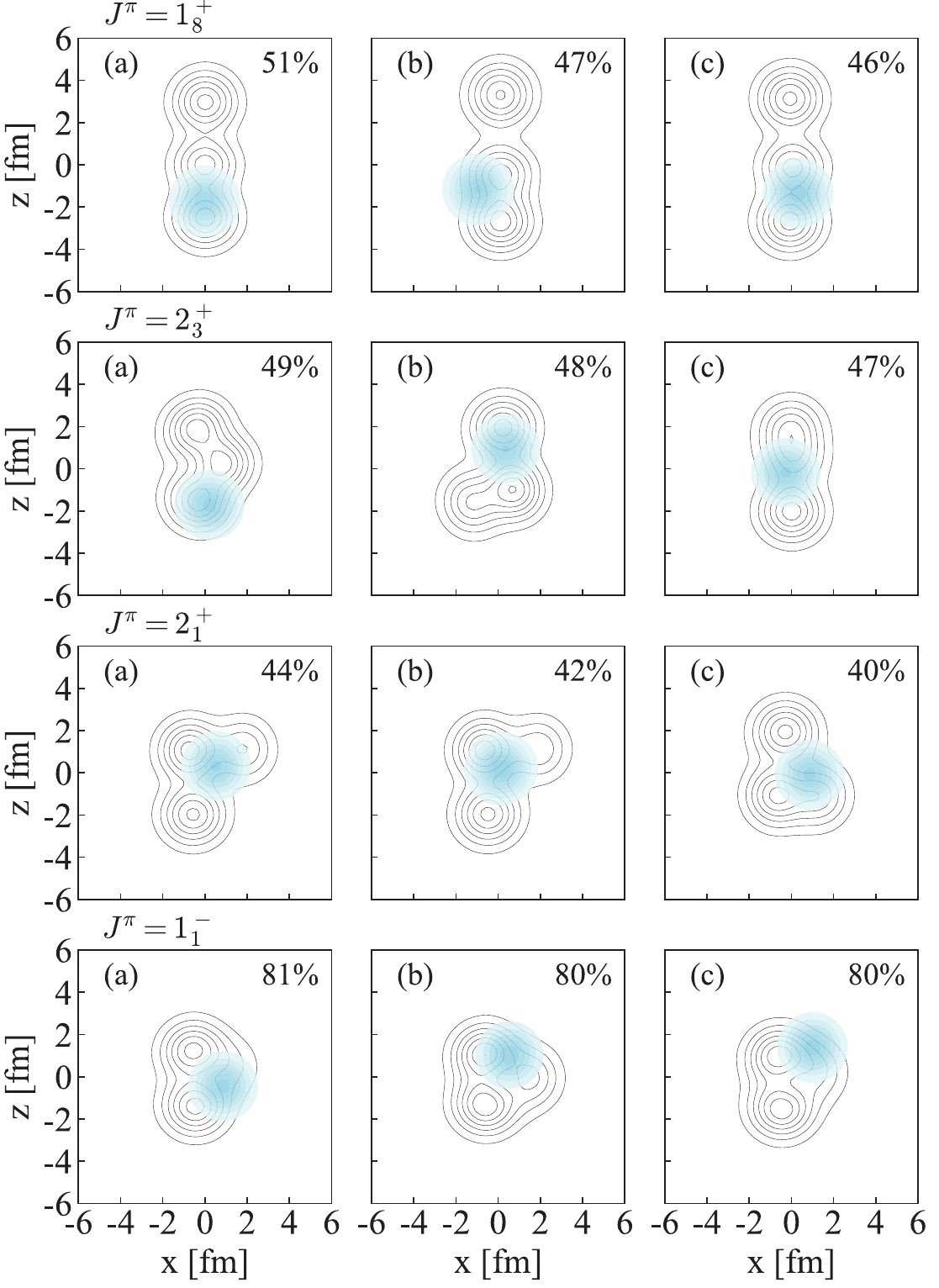} %插入图片，[]中设置图片大小，{}中是图片文件名
  \caption{The density profiles of main bases in $1_8^+$, $2_3^+$, $2_1^+$ and $1_1^-$ states of $_\Lambda^{12}\rm{B}$ where the solid lines show the nuclear distributions, blue portion shows the $\Lambda$ particle distribution. The number in each panel is the squared overlap $\mathcal{O}_i$ between each basis and total wave function.} %最终文档中希望显示的图片标题
  \label{density} %用于文内引用的标签
\end{figure}

Furthermore, we discuss the dynamic changes, focusing on the predominant configuration of the $s_\Lambda$ dominant states in $_\Lambda^{12}\rm{B}$ while parity coupling from different $\Lambda$ orbits and the shape coexistence of the core nucleus are considerable.
Fig.~\ref{density} shows the density profiles of the main bases with the largest squared overlap in several representative states, as well as the ground state $1_1^-$.
Overall, with the excitation energy increases, several distinct excitation modes of the $t$ cluster are produced in $_\Lambda^{12}\rm{B}$, comparing the significant compact structure of the ground state.
Most notably, the obtained $1_8^+$ state exhibits a linear-chain configuration. However, the linearly aligned $\alpha$-$t$-$\alpha$ configuration is more developed than the $\alpha$-$\alpha$-$t$ or $t$-$\alpha$-$\alpha$ configuration in $1/2_2^+$ ($^{11}\rm{B}$) \cite{zhou_2_2018}.  
The distribution of $t$ cluster is significantly contracted because of the shrinkage effect of $\Lambda$ hyperon.
In the $2_1^+$ state, the $t$ cluster shows a relatively compact distribution similar to $5/2_1^+$ core nucleus state because the $\Lambda$ hyperon prefers to be situated at the center of mass, which further attracts the $\alpha$ and $t$ clusters to form a compact configuration.
Moreover, despite the bent-arm structure of $^{11}\rm{B}$ remaining, a novel isosceles triangle configuration plays a dominant role in the $2_3^+$ state, which is not favored for the cluster states in $^{11}\rm{B}$ \cite{zhou_2_2018}. 
Panel (c) has a similar configuration in the y-z plane with panel (a) in the x-z plane.
The shape coexistence near peak $\#$7 is also induced by the strong attraction from the $\Lambda$ hyperon.
Additionally, the $\Lambda$-$\alpha$ correlation is also confirmed in the excited states, which has been revealed in the ground states of ${{}^{9-11}_{\Lambda} \rm{Be}}$ \cite{tian_lambda_2024}.
All the differences about these excited states indicate that the $\Lambda$ hyperon significantly affects the nucleon structures, impacting both the inter-cluster distances and the distribution of the $t$ cluster. 
With the increase of excitation energy, we can summarize that the structural evolution of $_\Lambda^{12}\rm{B}$ undergoes the four stages in Fig.~\ref{LNforce}. The initial shell-like structure can be evolved to the clustering structure. Around the $E_x = 12$ MeV, the $\Lambda$ particle is excited to the $p$ orbit and coupled with the liquid-like low-lying states of $^{11}\rm{B}$. By the further excitation ($E_x = 15$ MeV), the $\Lambda$ particle is de-excited and the $_\Lambda^{12}\rm{B}$ exhibits the linear-chain-like structure with the well developed clustering.

Now, we compare our results with the experimental peaks and provide the theoretical prediction within the cluster picture. Many experimental peaks with high excitation energies ($E_x = 9-13$ MeV) are observed in the pioneering $(e, e'K)$ hadronic reaction. Some of peaks ($\#$5, 6, 8) are explained as $p_\Lambda$ dominant states using shell-model calculations, while others ($\#$4, 7) remain unexplained \cite{iodice_high_2007,tang_experiments_2014}. 
Here, we first discuss possible candidates for these unexplained peaks, based on the level spacing and structural information of excited states.
Peak $\#$4 appears in a relatively ``clear" region below the $\Lambda$ breakup threshold. 
Therefore, considering its position, the $2_1^+$ ($5/2_1^+\otimes s_\Lambda$) state with a relatively compact configuration can be seen as a possible candidate for $\#$4. 
Another unexplained peak, $\#$7, is reported above the threshold experimentally. 
Nevertheless, no $s_\Lambda$ dominant states are found in this region.  
Given the absence of peak $\#$7 in shell-model calculations, the remaining possible states for this peak are $1_2^+$, $2_2^+$, and $2_3^+$, which exhibit the developed clustering structure, despite they are slightly underestimated in the current calculations.
On the other hand, peaks $\#$6 and $\#$8 are explained as ($3/2_1^-$, $1/2_1^-$) $\otimes$ $p_\Lambda$ states \cite{iodice_high_2007}, which are also supported by present calculations within cluster model. 
However, in present calculations, peak $\#$5 is more likely explained as the $1_1^+$, which is an $s_\Lambda$ dominant state, rather than a $p_\Lambda$ dominant state as predicted by the shell-model calculation \cite{iodice_high_2007}. 
Another contradiction is that the predicted cross-section of peak $\#$5 in the shell model calculation is only 26$\%$ of the experimentally observed value. 
Hence, it is very interesting and important to clarify the excitation mode of peak $\#$5 with theoretical cross-section predictions, considering its clustering nature and parity coupling effect.

\section{conclusion}\label{conclusion1}
In this Letter, we extend the microscopic cluster model to systematically investigate the positive-parity spectra of $_\Lambda^{12}\rm{B}$ hypernucleus, incorporating the Hyper-Brink wave function and a neural network-driven optimization framework. 
Through calculations and analyses, we successfully identify the excitation modes of each state, confirming the parity coupling effect in the $_\Lambda^{12}\rm{B}$ hypernucleus.
Overall, most states have a clustering nature and many novel configurations such as isosceles triangle and $\alpha$-$t$-$\alpha$ linear-chain configurations emerge due to the introduction of $\Lambda$ particle.
The $\Lambda$-$\alpha$ correlation is further confirmed in the $_\Lambda^{12}\rm{B}$ hypernucleus.
Furthermore, the structural evolution of $_\Lambda^{12}\rm{B}$ positive-parity states is summarized, ranging from compact shell-like structures to linear-chain configurations.
By comparing with experimental data, we identify peaks $\#$6 and $\#$8 as $p_{\Lambda}$ dominant states, consistent with shell-model predictions.
Based on the comprehension of $s_{\Lambda}$ dominant states and parity coupling within the cluster picture, we present the possible candidates of the unexplained ($\#$4, $\#$7) and controversial experimental peaks ($\#$5).
These findings highlight the importance of parity coupling and clustering in understanding the hypernuclear structure and the impact of strangeness. Overall, this work advances the theoretical understanding of hypernuclei and provides a basis for future experimental exploration of these exotic hypernuclear systems.

\section{acknowledge}
This work is supported by the National Natural Science Foundation of China
(Grants No.\ 12105141, 12205105, 12305123), by the Jiangsu Provincial Natural Science Foundation
(Grants No.\ BK20210277),  by the 2021 Jiangsu Shuangchuang (Mass Innovation and
Entrepreneurship) Talent Program (Grants No.\ JSSCBS20210169), by the JSPS KAKENHI Grant No.JP22K03643, by JST ERATO Grant No.JPMJER2304, and by the JSPS A3 Foresight Program, Japan. 
This work is partially supported by High Performance Computing Platform of Nanjing University of Aeronautics and Astronautics.

\section{Appendix}
To reveal the negative-parity component of the $\Lambda$ hyperon, we calculate the overlap of the total wave functions before and after applying a negative-parity projection operator for hyperon wave function.

Since there is no Pauli-blocking effect between the 
$\Lambda$ hyperon and nucleons, the total wave function (\ref{total_wave_function}) can be directly decomposed into a 
$\Lambda$ hyperon component and a nuclear component. We then define a partial negative-parity projection operator for the hyperon component.
\begin{equation}
  \begin{aligned}
    \hat{P}_{\Lambda}^{-} \Psi
    = \frac{\phi^{\Lambda}(\boldsymbol{r}, \boldsymbol{R}) - \hat{P}^r\phi^{\Lambda}(\boldsymbol{r}, \boldsymbol{R})}{2} \mathcal{A} \left\{ \prod_{k=1}^{n} \phi^{N}(\boldsymbol{r}_k, \boldsymbol{R}_k) \right\}
  \end{aligned}
\end{equation}
We impose the partial negative-parity projection to the state and calculate the overlap, which conveniently reveals the different levels of coupling within the state.
\begin{equation}
  \mathcal{\hat{O}} = \frac{
      | \langle \Psi^J | \hat{P}_{\Lambda}^{-} | \Psi^J \rangle |^2
  }{
      |\langle \Psi^J | \Psi^J \rangle \langle \Psi^J | \hat{P}_{\Lambda}^{-} | \Psi^J \rangle|
  }
\end{equation}

% \section{References}
\bibliographystyle{apsrev4-1}
\bibliography{paper-B12-bib}
\end{document}